\documentclass[twocolumn,eqsecnum,aps,showpacs]{revtex4}

\begin{document}   

\preprint{YITP-04-25}

\title{Topological-charge anomalies in supersymmetric theories 
with domain walls}
\author{K. Shizuya}
\affiliation{Yukawa Institute for Theoretical Physics\\
Kyoto University,~Kyoto 606-8502,~Japan }

\begin{abstract} 
Domain walls in 1+2 dimensions are studied to clarify some general
features of topological-charge anomalies in supersymmetric theories, 
by extensive use of a superfield supercurrent.
For domain walls quantum modifications of the supercharge algebra arise
not only from the short-distance anomaly but also from another source of
long-distance origin, induced spin in the domain-wall background, and the
latter dominates in the sum. A close look into the supersymmetric trace
identity,  which naturally accommodates the central-charge anomaly and 
its superpartners, shows an interesting consequence of the
improvement of the supercurrent: 
Via an improvement the anomaly in the central charge can be transferred
from induced spin in the fermion sector to an induced potential in the
boson sector.  This fact reveals a dual character, both fermionic and
bosonic, of the central-charge anomaly, which reflects the underlying
supersymmetry.
The one-loop superfield effective action is also 
constructed to verify the anomaly and BPS saturation of the
domain-wall spectrum.  
\end{abstract}


\maketitle

\section{Introduction} 

There is an interesting interplay between supersymmetry and topological
excitations. As Witten and Olive~\cite{WO} pointed out, in the
presence of topological excitations the supercharge algebra is modified
to include central charges and in certain supersymmetric theories the
spectrum of topological excitations which saturate  the
Bogomol'nyi-Prasad-Sommerfield (BPS) bound~\cite{B} classically is
determined exactly through the central charge. An argument based on
multiplet shortening for BPS-saturated excitations shows that 
saturation persists  beyond the classical level in many
cases~\cite{WO,W}.

It, however, remained somewhat obscure whether and how BPS saturation
could continue at the quantum level in some simple supersymmetric
theories where the excitation spectrum is affected by quantum corrections
and renormalization. 
In this connection, solitons (or kinks) in two-dimensional theories with
$N=1$ supersymmetry~\cite{VF} had long been examined by a number of
authors~\cite{Schon,KR,H,IM,CM,U,RN,NSNR,GJ,SVV,GLN,LSV}. It was
eventually shown by Shifman, Vainshtein and Voloshin~\cite{SVV} that the
central charge acquires a quantum anomaly so that, together with the
quantum correction to the kink mass, BPS saturation is maintained at the
quantum level.   Their analysis revealed the importance of enforcing
supersymmetry although actual calculations were made with component
fields. 

Fujikawa and van Nieuwenhuizen~\cite{FN} developed a superspace approach to
this problem and derived the central-charge anomaly by making 
a local supersymmetry transformation on the superfield.
Subsequently a superfield formulation of the central-charge anomaly  
was presented~\cite{KScc} by making
extensive use of a superfield supercurrent that places the
supercurrent, energy-momentum tensor and
topological current in a supermultiplet.

The purpose of this paper is to present a further study of
the central-charge anomaly, especially its origins and character, 
for domain walls in 1+2 dimensions, for which nontrivial BPS
saturation of the quantum spectrum has been reported~\cite{RNW,GRNW}. 
For solitons in two dimensions the central-charge anomaly derives
entirely from the superconformal anomaly (of short-distance origin). 
For domain walls in three dimensions, in contrast, 
quantum modifications of the supercharge algebra come not only from the
short-distance anomaly but also from another source of long-distance
origin, induced spin in the domain-wall background, and the latter
dominates in the sum.
We point out some interesting consequences of the "improvement" 
of the superfield supercurrent:
Via an improvement the central-charge operator changes its form while its
(physical) expectation value remains unchanged.  
One can thereby transform induced spin in the fermion sector into 
an induced potential in the boson sector.  
This fact reveals a dual character, both fermionic and bosonic, of the
central-charge anomaly, which reflects the underlying supersymmetry.

In Sec.~II we review some basic features of supersymmetric
theories with topological charges. 
In Sec.~III we calculate the one-loop effective action in superspace
and identify a possible anomaly in the central charge.
In Sec.~IV we introduce a superfield supercurrent, examine its
conservation law at the quantum level and determine the central-charge
anomaly and its superpartners.  
In Sec.~V we consider the improvement of the superfield supercurrent
and its effect on the supersymmetric trace identities. 
In Sec.~VI we study physical origins of the central-charge anomaly
and examine what happens in the case of extended
supersymmetry.
Section~VII is devoted to a summary and discussion.

\section{N=1 supersymmetry in three dimensions}

Let us first review some basic features of supersymmetric theories 
with topological excitations in three (or two) dimensions.
Consider the Wess-Zumino model~\cite{WZ} consisting of a real scalar field
$\phi$ and a real (Majorana) spinor field
$\psi_{\alpha} =(\psi_{1},\psi_{2})$, along with a real auxiliary field
$F$, described by the action $S = \int d^{3}x\,  {\cal L}$ and
\begin{equation}
 {\cal L} =  {1\over{2}}\, \{\bar{\psi} i {\slash\!\!\!\partial} \psi 
+ (\partial_{\mu}\phi)^{2} + F^{2}\}  + FW'(\phi) 
-  {1\over{2}}\, W''(\phi)\, \bar{\psi}\psi ,
\label{modelcomponent}
\end{equation}
with the Dirac matrices  (in a Majorana representation) 
\begin{equation}
\gamma^{0}=\sigma_{2}, \gamma^{1}=i\sigma_{3},  \gamma^{2}=i\sigma_{1}.
\end{equation}
Here $\bar{\psi}\equiv \psi\gamma^{0}= i(\psi_{2}, -\psi_{1})$ and 
$W'(\phi)=dW(\phi)/d\phi$, etc.
Eliminating the auxiliary field $F$ from ${\cal L}$,
 i.e., setting $F \rightarrow -W'(\phi)$, yields the potential term 
$-{1\over{2}}\, [W'(\phi)]^{2}$.
We suppose that the superpotential  $W(\phi)$ has more than one extrema
with $W'(\phi)=0$ so that the model supports topologically stable
excitations.  A simple choice~\cite{VF}
\begin{equation}
W(\phi)= {m^{2}\over{4\lambda}}\, \phi 
- {\lambda\over{3}}\, \phi^{3}
\label{WZmodel}
\end{equation}
 supports a classical static domain-wall solution 
\begin{equation}
\phi_{\rm DW} (x) =v\tanh(mx^{1}/2)
\label{DWsol}
\end{equation}
with $v=m/(2\lambda)$, 
uniform in $x^{2}$ and interpolating between the two distinct vacua 
with $\langle \phi \rangle_{\rm vac}=\pm v$ at
spatial infinities $x^{1}=\pm \infty$.  
The domain wall has a finite energy density (or surface tension)
\begin{equation}
M^{\rm cl}_{\rm DW}/L_{y} = m^{3}/(6\lambda^{2}),
\label{Mcl}
\end{equation}
where $L_{y}=\int dy$ denotes the length in the $x^{2}\equiv y$ direction.
In two dimensions the same solution~(\ref{DWsol}) describes a static
kink~\cite{VF} with energy 
$M^{\rm cl}_{\rm kink} = m^{3}/(6\lambda^{2})$.
The super-sine-Gordon model with $W(\Phi) = mv^{2} \sin (\Phi/v)$
also supports analogous domain walls and solitons.

The action $S=\int d^{3}x\, {\cal L}$ is invariant under
supersymmetry transformations
\begin{eqnarray}
\delta \phi (x) &=& \bar{\xi}\psi (x), 
\delta \psi_{\alpha} (x) = -i (\gamma^{\mu}\xi)_{\alpha}
\partial_{\mu}\phi(x) + \xi_{\alpha} F(x), \nonumber\\
\delta F(x) &=& -i \bar{\xi}\gamma^{\mu} \partial_{\mu}\psi(x).
\label{susycomponent}
\end{eqnarray}
where $\xi_{\alpha}=(\xi_{1}, \xi_{1})$ is a two-component Grassmann
number; $\bar{\xi}= \xi\gamma^{0}$
and $\bar{\xi}\psi= \bar{\xi}_{\alpha}\psi_{\alpha}$.
The associated Noether supercurrent is written as
\begin{equation}
J^{\mu}_{\alpha} =
(\partial_{\nu}\phi)(\gamma^{\nu}\gamma^{\mu} \psi)_{\alpha}
-iF\, (\gamma^{\mu} \psi)_{\alpha}.
\label{Jmualpha}
\end{equation}
The conserved supercharges
\begin{equation}
Q_{\alpha} = \int d^{2}x\, J^{0}_{\alpha}
\end{equation}
generate, within the canonical formalism,
the transformation law of the supercurrent
\begin{equation}
i[\bar{\xi}_{\beta} Q_{\beta}, J^{\mu}_{\alpha}]
= -2i(\gamma_{\lambda}\xi)_{\alpha} (T^{\mu \lambda} 
+\epsilon^{\mu\lambda\nu}F \partial_{\nu}\phi ),
\label{deltaJ}
\end{equation}
with the canonical energy-momentum tensor 
\begin{equation}
T^{\mu \lambda} = {i\over{2}}\, \bar{\psi} 
\gamma^{\mu}\partial^{\lambda }\psi
+ \partial^{\mu}\phi\partial^{\lambda}\phi 
- {1\over{2}}\, g^{\mu \lambda}\{(\partial_{\nu}\phi)^{2} 
- F^{2}\}  
\label{Thetamunu}
\end{equation}
and the topological current
\begin{equation}
\epsilon^{\mu\lambda\nu}F \partial_{\nu}\phi 
= -\epsilon^{\mu\lambda\nu} \partial_{\nu}W(\phi).
\label{zetanaive}
\end{equation}
Here $F$ stands for $-W'(\phi)$ owing to the equation of
motion $\delta S/\delta F = F + W'(\phi)\rightarrow 0$. In deriving
Eq.~(\ref{deltaJ}) use has been made of the matrix identity specific to
1+2 dimensions,
\begin{equation}
\gamma^{\mu}\gamma^{\nu}
= g^{\mu \nu} - i\epsilon^{\mu\nu\lambda} \gamma_{\lambda}
\end{equation} 
with $\epsilon^{012}=1$.

Let us note that the energy-momentum tensor
$T^{\mu\lambda}$ has a portion antisymmetric in $(\mu, \lambda)$, 
\begin{equation}
T^{\mu \lambda}_{\rm asym} 
= -{1\over{8}}\,  \epsilon^{\mu \lambda\nu}
\{\partial_{\nu}(\bar{\psi}\psi)  + 2X_{\nu} \}.
\end{equation}
[In two dimensions 
$T^{\mu \lambda}_{\rm asym} \propto \epsilon^{\mu \lambda} 
\bar{\psi}\gamma^{0}\gamma^{1}(\delta S/\delta \bar{\psi})$ 
vanishes (at the quantum level~\cite{KScc}).]
Here 
\begin{equation}
X^{\nu} = -\bar{\psi}\,\gamma^{\nu}\gamma^{\rho}\partial_{\rho}\psi
=i\bar{\psi}\, \gamma^{\nu}(\delta S/\delta \bar{\psi})
\label{Xmu}
\end{equation}
is proportional to the equation of motion 
$\delta S/\delta \bar{\psi} = i\gamma^{\mu}\partial_{\mu}\psi 
- W''\psi  \rightarrow 0$ and vanishes.
It is thus natural to isolate the symmetric part 
$\Theta^{\mu\lambda} \equiv T^{\mu \lambda}_{\rm sym}$ and
to regard $T^{\mu \lambda}_{\rm asym}$ as part of the
(canonical) topological current 
\begin{equation}
\zeta_{\rm can}^{\mu\lambda}
=\epsilon^{\mu\lambda\nu}
\{ F\partial_{\nu}\phi - (1/8)\, \partial_{\nu}(\bar{\psi}\psi) \},
\label{topolcurrent}
\end{equation}
which may be  rewritten as $-\epsilon^{\mu\lambda\nu}\partial_{\nu}
\{ W(\phi) + (1/8)\,\bar{\psi}\psi \}$.

From Eq.~(\ref{deltaJ}) follows the supercharge algebra
\begin{equation}
\{Q_{\alpha}, \bar{Q}_{\beta}\} 
= 2(\gamma_{\lambda})_{\alpha\beta}\, (P^{\lambda} +Z^{\lambda})
\label{superchargealgebra}
\end{equation}
where $P^{\lambda}=\int d^{2}{\bf x} \Theta^{0\lambda}$ is the total
energy and momentum. The central charge 
\begin{equation}
Z^{\lambda}=\int d^{2}{\bf x}\, \zeta^{0\lambda}_{\rm can},
\label{Znaive}
\end{equation}
for the classical domain-wall configuration 
$\phi_{DW}(x) \rightarrow \pm v$ and 
$\psi_{\alpha}(x) \rightarrow 0$  as $x^{1}\rightarrow \pm \infty$, 
reads $Z^{1}=0$ and
\begin{equation}
 Z^{2}/ L_{y} = \Big[ W(\phi)\Big]_{x^{1}= -\infty}^{x^{1}=\infty}
= m^{3}/(6\lambda^{2}).
\label{expZtwo}
\end{equation}
The $N=1$ superalgebra thus gets centrally extended in the presence of
domain walls (as well as solitons in two dimensions)~\cite{WO}.

The Wess-Zumino model~(\ref{modelcomponent}) is neatly rephrased using
the superfield formalism~\cite{WB}.
The structure of $N=1$ supersymmetry is formally the same for two and
three dimensions, with points
$z=(x^{\mu},\theta_{\alpha})$ in $N=1$ superspace labeled by
spacetime coordinates $x^{\mu}$ and two Majorana coordinates
$\theta_{\alpha}=(\theta_{1},\theta_{2})$.  
The supermultiplet nature of the fields 
$(\phi, \psi_{\alpha}, F)$ is encoded in a real superfield, 
\begin{equation}
\Phi (z) \equiv \Phi (x,\theta) = \phi (x) + \bar{\theta}\psi (x) 
+ {1\over{2}}\,\bar{\theta}\theta\, F(x),
\end{equation}
where $\bar{\theta}\equiv \theta\gamma^{0} 
= i(\theta_{2}, -\theta_{1})$ and  
$\bar{\theta}\theta=\bar{\theta}_{\alpha}\theta_{\alpha}=
-2i\theta_{1}\theta_{2}$.
Under translations
$x^{\mu} \rightarrow x^{\mu} - i\bar{\theta} \gamma^{\mu} \xi$
and $\theta_{\alpha} \rightarrow \theta_{\alpha} +\xi_{\alpha}$
in superspace, the component fields undergo the supersymmetry
transformations~(\ref{susycomponent}).

The action $S = \int d^{3}x\, {\cal L}$ is cast in a superfield
form~\cite{VF}
\begin{equation}
 S[\Phi] = \int d^{5}z\, \left\{ {1\over{4}}\,
(\bar{D}_{\alpha}\Phi)\, D_{\alpha}\Phi + W(\Phi)
\right\}
\label{sfaction}
\end{equation}
with $d^{5}z= d^{3}xd^{2}\theta$ and $\int d^{2}\theta\,
{1\over{2}}\,\bar{\theta}\theta =1$. Here the spinor derivatives 
\begin{equation}
D_{\alpha}= \partial/\partial \bar{\theta}_{\alpha}
- (\slash\!\!\! p\, \theta)_{\alpha}
\end{equation}
and $\bar{D}_{\alpha}\equiv D_{\beta}(\gamma^{0})_{\beta \alpha}$, 
with
$\slash\!\!\! p \equiv \gamma^{\mu}p_{\mu}$ and
$p_{\mu}=i\partial_{\mu}$, 
obey 
\begin{equation}
\{D_{\alpha}, D_{\beta}\} 
= 2\, (\slash\!\!\! p \gamma^{0})_{\alpha\beta};
\end{equation}
see Ref.[\onlinecite{KScc}] for some formulas involving
$D_{\alpha}$.

The superalgebra~(\ref{superchargealgebra}) has an important
consequence.  For the supercharge $Q_{2}$, in particular, it reads
\begin{equation}
(Q_{2})^{2} = P^{0} -P^{2}-Z^{2}.
\end{equation}
The classical domain-wall solution~(\ref{DWsol}) (giving 
$P^{2}=0$)  obeys the  first-order equation
$(\partial/\partial x^{1}) \phi = W'(\phi)$, and is BPS
saturated~\cite{WO} in the sense that it is inert 
under the action of $Q_{2}$, 
\begin{equation}
Q_{2} |{\rm DW}\rangle  = (P^{0} - Z^{2})|{\rm DW}\rangle =0.
\label{BPSsaturation}
\end{equation}
The supercharge $Q_{1}$, on the other hand, acts nontrivially.
The BPS saturation~(\ref{BPSsaturation}) thus implies that 
the domain-wall tension $\langle P^{0} \rangle$ is given by the central
charge $\langle Z^{2} \rangle$ exactly; 
here $\langle \cdots \rangle$ stands for the expectation
value $\langle {\rm DW}| \cdots |{\rm DW} \rangle$ for short.  
It is clear from Eqs.~(\ref{Mcl}) and~(\ref{expZtwo}) that 
this equality holds at the classical level.

The BPS saturation $Q_{2} |{\rm DW}\rangle  =0$ and 
the resulting equality 
$\langle P^{0} \rangle = \langle Z^{2} \rangle$, once established
classically, generally persist at the quantum level. 
This follows from multiplet shortening for BPS-saturated excitations in
many cases~\cite{WO}.   Solitons in two dimensions and the domain wall
under consideration also belong to short one-dimensional representations,
preserving only half of the original supersymmetry~\cite{SVV,LSV}.

Some remarks are in order here.  To validate the formal reasoning based on
the superalgebra, one has to preserve supersymmetry in actual
calculations.   This is most naturally achieved by use of superfields, as
we shall verify later.
As a result, the supercharge algebra~(\ref{superchargealgebra})  holds as
it is at the quantum level  
[although the charges $(Q_{\alpha}, P^{\lambda}, Z^{\lambda})$ may deviate
from their classical expressions;  see, e.g., Eq.~(\ref{zetaanom})]. 
The BPS saturation~(\ref{BPSsaturation}) implies
that the domain-wall superfield has the form~\cite{SVV}
\begin{equation}
\Phi_{\rm DW}(z)
= \phi_{\rm DW} \Big( x^{1}  -{1\over{2}}\,\bar{\theta}\theta \Big),
\end{equation}
which thus relates the domain-wall background field $\phi_{\rm
DW}(x^{1})$ and the associated $F$ component
so that
\begin{equation}
\partial_{1}\phi_{\rm DW}(x^{1})
= - F_{\rm DW}(x^{1}).
\end{equation}
Note that the action of $Q_{2}$ (or supertranslations with
$\xi_{1}$) preserves the interval $x^{1}
-{1\over{2}}\,\bar{\theta}\theta$ and hence $\Phi_{\rm DW}(z)$ as well.

To fix the functional form of $\phi_{\rm DW}(x^{1})$ one may
use the effective action in superspace. 
Suppose we have calculated the effective action~\cite{CW}
$\Gamma[\Phi_{c}]$ as a functional of the classical field 
$\Phi_{\rm c}(z)$ [$= \langle \Phi (z) \rangle$ in the presence of 
a classical source $J_{\Phi}(z)$] in a loop-wise expansion.  
It is  a sum of the classical action~(\ref{sfaction}) and 
loop corrections, 
$\Gamma[\Phi_{c}] = S[\Phi_{c}] + \Gamma_{\rm loop}[\Phi_{\rm c}]$, and 
the associated classical equation of motion 
$\delta \Gamma[\Phi_{c}]/\delta\Phi_{\rm c}(z) =0$ 
governs the quantum dynamics of $\Phi_{\rm c}(z)$.  
The key fact~\cite{KScc} is that this superfield equation 
$\delta \Gamma[\Phi_{c}]/\delta\Phi_{\rm c}(z) =0$ directly
turns into the BPS equation for $\Phi_{\rm DW}(z)$, 
on substitution $\Phi_{\rm c}(z) \rightarrow  \Phi_{\rm DW}(z)$ 
(and on noting that $D_{1}\Phi_{\rm DW}=0$ and
$\bar{D}D\Phi_{\rm DW}= 2\partial_{1}\Phi_{\rm DW}$).  
The superspace effective action thus accommodates BPS-saturated
excitations quite naturally.

\section{Superspace effective action}

In this section we calculate the effective action and identify 
a possible anomaly in the central charge.
We use the background-field method~\cite{CW} and expand $\Phi$ around
the classical field $\Phi_{\rm c}$,
$\Phi (z) = \Phi_{\rm c}(z) + \chi(z)$.
The quantum fluctuation $\chi$ at the one-loop level is governed by the
action $\int d^{5}z\,  
{1\over{2}}\, \chi {\cal D} \chi$ with the superspace operator 
\begin{equation}
{\cal D} = -{1\over{2}}\bar{D}D + W''(\Phi_{\rm c}).
\end{equation}
The associated $\chi$ propagator is given by $i/{\cal D}$,
which we regularize in a supersymmetric way as 
\begin{equation}
\langle \chi (z)\,\chi (z')\rangle^{\rm reg} = \langle z|
{i\over{\cal D}}\, e^{\tau {\cal D}^{2}} |z'\rangle,
\label{regpropagator}
\end{equation}
with $\tau \rightarrow 0_{+}$ in the ultraviolet (UV) regulator 
$e^{\tau {\cal D}^{2}}$.

One can evaluate the $\chi$ propagator by expanding it in powers of
$D_{\alpha}$ acting on $M \equiv W''(\Phi_{\rm c})$. 
The calculation is essentially the same as in the two-dimensional (2d)
kink case;  one may simply evaluate Eq.~(C2) of Ref.~\cite{KScc} in 3
dimensions.  To $O(D^{2})$  the result is 
\begin{equation}
\langle \chi (z)\,  \chi (z)\rangle 
=  2\kappa  -{|M|\over{4\pi}}  - {\bar{D}DM \over{16\pi|M|}}
+ {(\bar{D}M) DM \over{32\pi M|M|}},
\label{chipropagator}
\end{equation}
where $\kappa \equiv 1/(8\pi \sqrt{\pi \tau})$ is UV-cutoff dependent.
Integrating this with respect to $\Phi_{\rm c}$, as done
earlier~\cite{KScc}, then yields the one-loop effective action to
$O(D^{2})$,
\begin{equation}
\Gamma_{1}[\Phi_{\rm c}]
=  \int d^{5}z\,  \Big[ \kappa M - {M|M|\over{16\pi}} 
+ {(\bar{D}_{\alpha}M) D_{\alpha}M \over{64\pi |M|}} \Big].
\end{equation}

The $O(D^{0})$  terms in the total one-loop effective action 
$\Gamma [\Phi_{c}] = S[\Phi_{c}] + \Gamma_{1}[\Phi_{\rm c}]$ now read 
\begin{equation}
U_{\rm eff}(\Phi_{\rm c}) =  W(\Phi_{\rm c})  
+ \kappa W''_{\rm c}
-{1\over{16\pi}}\, W''_{\rm c}|W''_{\rm c}| ,
\label{Ueff}
\end{equation}
where $W''_{\rm c}\equiv M= W''(\Phi_{\rm c})$.  
Rewriting it in favor of the expectation value 
$\langle W(\Phi) \rangle = W(\Phi_{\rm c}) 
+ {1\over{2}}\, W''_{\rm c}\, \langle \chi \chi \rangle 
+ \cdots$
yields
\begin{equation}
U_{\rm eff}(\Phi_{\rm c}) 
=  \langle W(\Phi) \rangle 
+ {1\over{16\pi}}\, W''_{\rm c}|W''_{\rm c}|.
\label{UeffW}
\end{equation}
This shows that the superpotential deviates from the classical
superpotential (operator) $W(\Phi)$  by
$(1/16 \pi)\, W''(\Phi) |W''(\Phi)|$ at the one-loop level, suggesting 
a quantum anomaly in the central charge.
[The identification~(\ref{UeffW}) is meant to $O(D^{0})$. Interestingly,
its right-hand side agrees with 
$\Gamma_{1}[\Phi_{\rm c}]$ to $O(D^{2})$, since the difference 
$\sim (1/32\pi)\,  \bar{D}D|W''_{\rm c}|$ vanishes
under $\int d^{5}z$.]

The UV-divergent term $\kappa W''_{\rm c}$ in 
$U_{\rm eff}(\Phi_{\rm c})$ can be eliminated by mass renormalization. 
To this end let $m_{\rm r}$ be a finite mass scale and set
$m^{2}=m_{\rm r}^{2}  + \delta m^{2}$ in $U_{\rm eff}(\Phi_{\rm c})$.
A convenient choice for the mass counterterm is 
\begin{equation}
\delta m^{2} = 8\lambda^{2} \kappa,
\end{equation}
the net effect of which is to 
set $m\rightarrow m_{\rm r}$ and $\kappa\rightarrow 0$ in 
$U_{\rm eff}(\Phi_{\rm c})$. 

The effective action $\Gamma [\Phi_{c}]$ to $O(D^{2})$ governs the
asymptotic $({\bf x} \rightarrow \pm \infty)$ characteristics of the
domain-wall state, which are sufficient for determining the
central charge and for verifying BPS saturation.  
Retaining only the bosonic components
$\phi_{\rm c}(x)$ and $F_{\rm c}(x)$ of $\Phi_{\rm c}(z)$ 
in $\Gamma [\Phi_{c}]$, one obtains the Lagrangian for the
static wall,
\begin{eqnarray}
 {\cal L}_{\rm stat}
&=&  -{1\over{2}}\,\Big\{\sqrt{\alpha}\, \partial_{1}\phi_{\rm c}
\mp {1\over{\sqrt{\alpha}}}\, U'_{\rm eff}(\phi_{\rm c})\Big\}^{2} 
\nonumber\\
&& \mp\, \partial_{1} U_{\rm eff}(\phi_{\rm c}) 
-{1\over{2}}\,\alpha\, (\partial_{2}\phi_{\rm c})^{2},
\end{eqnarray}
with 
$\alpha (\phi_{\rm c}) = 1 + (1/16\pi)\,
\{W'''(\phi_{\rm c})\}^{2}/|W''(\phi_{\rm c})|$.
This leads to the BPS equation for the domain wall,
\begin{equation}
\partial_{1}\phi_{\rm c}
= -F_{\rm c}= (1/\alpha)\, U'_{\rm eff}(\phi_{\rm c}),
\partial_{2}\phi_{\rm c} = 0,
\label{BPSeq}
\end{equation}
with the asymptotic values of $\phi_{\rm c}$ at $x^{1}=\pm \infty$ now
determined from $U'_{\rm eff}(\phi_{\rm c})\equiv  
dU_{\rm eff}(\phi_{\rm c})/ d\phi_{\rm c} = 0$. 
[The superfield equation 
$\delta \Gamma[\Phi_{c}]/\delta\Phi_{\rm c}(z) =0$ 
also leads to the same BPS equation.]   
The central charge 
$Z_{\rm c} = \int d^{2}{\bf x}\, \partial_{1} U_{\rm eff}(\phi_{\rm c})$
then gives the surface tension 
\begin{equation}
M_{\rm DW}/L_{y} = Z_{\rm c}/L_{y}
= {m_{\rm r}^{3}\over{6\lambda^{2}}} -{m^{2}_{\rm r}\over{8\pi}}
\label{dwtension}
\end{equation}
at $\phi_{\rm c}(x^{1}=\infty) = m_{\rm r}/(2\lambda) -\lambda/(4\pi)$, 
in agreement with earlier results~\cite{RNW}. 
Here the quantum correction $m^{2}_{\rm r}/(8\pi)$ derives 
from $(1/16\pi)\, W''_{\rm c}|W''_{\rm c}|$ 
in $U_{\rm eff}(\Phi_{\rm c})$.  Note that it is the potential 
$U_{\rm eff}(\Phi_{\rm c})$ that should be minimized, rather than 
the operator potential
$W(\Phi) + (1/16\pi)\, W''(\Phi)|W''(\Phi)|$ 
which, upon minimization, leads to a (divergent) unrenormalized 
expression with a quantum correction of the wrong sign.

Similarly, the super-sine-Gordon model with $W(\Phi)= m v^{2} \sin
(\Phi/v)$ leads to the domain-wall surface tension 
\begin{equation}
M_{\rm DW}/L_{y} = 2m_{\rm r} v^{2} -(m^{2}_{\rm r}/8\pi), 
\end{equation}
upon setting $m=m_{\rm r}  + \delta m$ and 
$\delta m = \kappa\,  m_{\rm r}/v^{2}$.

\section{Superfield supercurrent}

In this section we study possible quantum modifications of
the supercharge algebra.
The first step is to make a proper choice of conserved symmetry
currents.
This is not an easy step if one notes that there is 
some arbitrariness in defining  currents 
(such as $J^{\mu}_{\alpha}$, $\Theta^{\mu \lambda}$, etc.) 
in supersymmetric theories: 
One may use either the auxiliary field $F$ or its (classical) equivalent
$-W'(\phi)$ to form currents but such possible choices are not
necessarily the same at the quantum level, as we shall see soon.

Fortunately, in the present case one may simply adopt a superfield
supercurrent used in the 2d kink case~\cite{KScc}, 
which (now adapted to 1+2 dimensions) reads
\begin{equation}
{\cal V}^{\mu}_{\alpha} 
= -i(D_{\alpha}\bar{D}_{\lambda}\Phi)\, 
(\gamma^{\mu})_{\lambda\beta}\,D_{\beta}\Phi.
\label{sfcurrent}
\end{equation}
This current is a real spinor-vector superfield and places 
the supercurrent $J^{\mu}_{\alpha}$, 
energy-momentum tensor $T^{\mu \lambda}$ and topological current 
in a supermultiplet, as seen from the component expression
\begin{equation}
{\cal V}^{\mu}_{\alpha}
= J^{\mu}_{\alpha} -2i(\gamma_{\lambda}\theta)_{\alpha} 
(T^{\mu \lambda} + \epsilon^{\mu \lambda\nu}F\partial_{\nu}\phi)
+\theta_{\alpha}X^{\mu} 
+{1\over{2}}\, \bar{\theta}\theta f^{\mu}_{\alpha}. 
\label{sfcurrentcomponent}
\end{equation}
Here $J^{\mu}_{\alpha}$ and $T^{\mu\lambda}$ are defined  by
Eqs.~(\ref{Jmualpha}) and~(\ref{Thetamunu}) [with the auxiliary field
$F$ not identified with $-W'(\phi)$], respectively;
$X^{\mu} =i\bar{\psi}\, \gamma^{\mu}(\delta S/ \delta \bar{\psi})$ 
as defined in Eq.~(\ref{Xmu}).   We refer to one more current
$f^{\mu}_{\alpha}$  somewhat later.
This current ${\cal V}^{\mu}_{\alpha}$ obeys a conservation
law of the form
\begin{equation}
\partial_{\mu}{\cal V}^{\mu}_{\alpha} = 
(D_{\alpha}\bar{D}_{\beta}\Phi)\,D_{\beta}{\delta S\over{\delta \Phi}}
-(D_{\alpha}\bar{D}_{\beta}{\delta S\over{\delta \Phi}})\,
D_{\beta}\Phi,
\label{dValpha}
\end{equation}
where 
\begin{equation}
{\delta S\over{\delta \Phi}} 
= - {1\over{2}}\,\bar{D}D\Phi + W'(\Phi)
\label{dSdPhi}
\end{equation}
is an identity.
One would think that current conservation $\partial_{\mu }{\cal
V}^{\mu}_{\alpha} = 0$  simply follows from the equation of
motion $\delta S/\delta\Phi=0$.  Care is required here, however.
In general, while the equations of motion hold by themselves,
operator products of the form (equations of motion)$\times$(fields)
are potentially singular and, when properly regulated, may not vanish.  
Indeed, in Fujikawa's method~\cite{F,Ftwo}
all known anomalies arise from regularized Jacobian factors which take
precisely such form.  
One therefore has to keep track of such potentially anomalous products
to determine the conservation laws at the quantum level; 
see Ref.~\cite{KS} for an early study of the superconformal anomaly
along this line.

Here we quote only some general features of the potentially anomalous
products, studied earlier~\cite{KScc}.   
Consider a product of the form
\begin{equation}
\{\Omega \Phi (z) \}\, \Xi {\delta S\over{\delta \Phi (z)}} ,
\label{XiLambda}
\end{equation}
where $\Omega$ and $\Xi$ may involve operators $D_{\alpha}$ and
$\partial_{\mu}$.
One can  evaluate it using the regularized
propagator~(\ref{regpropagator}) at the one-loop level.
The key result is that the regularized products enjoy the
reciprocal property
\begin{equation}
(\Omega \Phi)\, \Xi {\delta S\over{\delta \Phi}}
= \pm\, (\Xi \Phi)\, \Omega {\delta S\over{\delta \Phi}},
\label{keyformula}
\end{equation}
where the minus sign applies only when both $\Omega$ and
$\Xi$ are Grassmann-odd.
An immediate consequence of this formula and 
Eq.~(\ref{dValpha}) is the conservation of the supercurrent
$\partial_{\mu}{\cal V}^{\mu}_{\alpha}=0$ at the quantum level. 
The simplest anomalous product we shall use later is
\begin{equation}
\Phi {\delta S\over{\delta \Phi}} = -2\kappa W''(\Phi)
\label{phidsdphi}
\end{equation}
at the one-loop level, with 
$\kappa \equiv 1/(8\pi \sqrt{\pi \tau})$ as before; for a derivation
one may evaluate Eq.~(B2) of Ref.~\cite{KScc} in three dimensions.

The highest component in ${\cal V}^{\mu}_{\alpha}$ is written as
\begin{equation}
f^{\mu}_{\alpha} 
= -2\epsilon^{\mu\lambda\nu}\partial_{\lambda}(iF\gamma_{\nu}\psi
-\phi\partial_{\nu}\psi) + r^{\mu}_{\alpha},
\label{highestcurrent}
\end{equation}
where $r^{\mu}_{\alpha}$ collectively stands for potentially anomalous
products which take essentially the same form as in the 2d kink
case.  One can show quite generally, using the
formula~(\ref{keyformula}), that $X^{\mu}=r^{\mu}=0$ at the quantum
level~\cite{KScc}.
Correspondingly, the associated spinor charge vanishes
\begin{equation}
\int_{-\infty}^{\infty} d^{2}{\bf x}\, f^{0}_{\alpha} = 0,
\label{fmualpha}
\end{equation} 
as long as the spinor field $\psi_{\alpha} \rightarrow 0$ for
$x^{1}\rightarrow \pm \infty$ while all the fields, like the
classical domain-wall configuration $\phi_{DW}(x)$, are  uniform for
$x^{2}\rightarrow \pm \infty$ 
[so that $\phi_{i}|_{x^{2}=\infty} =\phi_{i}|_{x^{2}=-\infty}$ 
with $\phi_{i} =(\phi, \psi_{\alpha}, F)$].
Hence only $Q_{\alpha}$, $P^{\lambda}$ and $Z^{\lambda}$ form an
irreducible supermultiplet, yielding a conserved-charge superfield
\begin{equation}
\int_{-\infty}^{\infty} d^{2}{\bf x} {\cal V}^{0}_{\alpha}
= Q_{\alpha} -2i(\gamma_{\lambda}\theta)_{\alpha} 
(P^{\lambda} + Z^{\lambda}) ,
\label{sfcharge}
\end{equation}
which, upon supertranslations, correctly reproduces the supercharge
algebra~(\ref{superchargealgebra}).

While anomalous products have left the conservation law 
$\partial_{\mu}{\cal V}^{\mu}_{\alpha}=0$ intact, 
they cause some changes in the component currents of 
${\cal V}^{\mu}_{\alpha}$.
Consider, e.g., the topological current
$\zeta^{\mu\lambda}_{\rm can}$ defined by
Eq.~(\ref{topolcurrent})  [with $X_{\nu}=0$] and rewrite the first term 
as $F\partial_{\nu}\phi = -\partial_{\nu}W
+ (\delta S/\delta F)\partial_{\nu}\phi$, using 
$\delta S/\delta F = F + W'$.  The key formula~(\ref{keyformula})
then implies that 
$(\delta S/\delta F) \partial_{\nu}\phi 
=(1/2)\, \partial_{\nu}( \phi\, \delta S/\delta F)$,
with the anomalous product 
$\phi\, (\delta S/\delta F) = -2\kappa W''(\phi)$ 
read~\cite{fndSdphi} from the superfield product~(\ref{phidsdphi}).
In effect, $F$ multiplied with $\partial\phi$ acts 
like $-(W' + \kappa W''')$ at the quantum level;
the auxiliary field $F$ thus changes
its role in composite operators.  
As a result $\zeta^{\mu\lambda}_{\rm can}$ deviates from the classical
expression~(\ref{topolcurrent}) $\sim  O(\hbar^{-1})$ by
$\kappa W''\sim O(\hbar^{0})$,
\begin{equation}
\zeta^{\mu\lambda}_{\rm can}
= -\epsilon^{\mu\lambda\nu}\partial_{\nu}
\Big\{ W(\phi) + {1\over{8}}\,\bar{\psi}\psi 
+\kappa W''(\phi) \Big\}.
\label{zetaanom}
\end{equation}
Note, however, that $\zeta^{\mu\lambda}_{\rm can}$, on eliminating $F$,
is only apparently modified; its very definition~(\ref{topolcurrent})
with $F$ is left intact. 
Analogous apparent modifications take place in $J_{\alpha}^{\mu}$ and
$\Theta^{\mu\nu}$ as well.  This is the general manner how the symmetry
currents in supersymmetric theories accommodate quantum anomalies
while leaving their supermultiplet structure and conservation laws
untouched, as observed earlier in the 2d kink case~\cite{KScc}.

The central charge 
$\langle Z^{\lambda}\rangle 
= \int d^{2}{\bf x} \langle \zeta^{0\lambda}_{\rm can} \rangle$ 
is now related to the operators $W(\phi), W''(\phi)$ and
$\bar{\psi}\psi$ at spatial infinities ${\bf x} \rightarrow \pm \infty$.
The composite operators $\bar{\psi}\psi$ and $\phi^{2}$, 
in general, become nonvanishing in the presence of
a classical field $\phi_{\rm c}(x)$, as seen from 
$\langle \chi \chi \rangle$ in Eq.~(\ref{chipropagator}). 
Actually, using the relation 
$(1/2)\bar{D}D\Phi^{2} = \Phi \bar{D}D\Phi +(\bar{D}\Phi) D\Phi$, 
one can relate $\langle \bar{\psi}\psi \rangle \sim 
\langle \bar{D}\chi(z)\, D\chi(z) \rangle$ to 
$\langle \chi \chi \rangle$,
\begin{eqnarray}
 \langle \bar{D}\chi D\chi  \rangle
&=& \{ {1\over{2}}\bar{D}D - 2W''_{\rm c}\} \langle \chi \chi  \rangle 
+2 \langle \Phi \delta S/\delta \Phi \rangle \nonumber\\
&=&  -8\kappa W''_{\rm c} 
+  {1\over{2\pi}}  W''_{\rm c}|W''_{\rm c}| + O(D^{2}).\ \ \
\label{DchiDchi}
\end{eqnarray}
In forming $\langle Z^{\lambda}\rangle$, the divergent term in $\langle
\bar{\psi}\psi \rangle$ and  the short-distance anomaly $\kappa W''$
combine to cancel so that 
\begin{equation}
{1\over{8}} \langle \bar{\psi}\psi\rangle 
+ \kappa \langle W''(\phi)\rangle  
= {1\over{16\pi}}  W''_{\rm c}|W''_{\rm c}| +\cdots,
\label{psibarpsi}
\end{equation}
in confirmation of $U_{\rm eff}(\Phi_{\rm c})$ in
Eq.~(\ref{UeffW}) and hence the surface tension~(\ref{dwtension}).

\section{Improvement and trace identities}

In this section we examine the central-charge anomaly 
in the light of superconformal symmetry. 
The supercurrent ${\cal V}^{\mu}_{\alpha}$ is composed of 
super-Poincare currents and is also used~\cite{FZ} to construct the
superconformal currents. 
As seen from the conservation law $\partial_{\mu} ({\slash \!\!\! x}
{\cal V}^{\mu}) =\gamma_{\mu}{\cal V}^{\mu}$ of the first-moment
current, in particular, explicit breaking to superconformal symmetry is
characterized by the quantity
$\gamma_{\mu}{\cal V}^{\mu}$.
Writing 
$i(\gamma_{\mu}{\cal V}^{\mu})_{\alpha} = 2 (\bar{D}D\Phi)D_{\alpha}\Phi
- (D_{\alpha}\bar{D}_{\beta}\Phi) D_{\beta}\Phi$
and isolating a term involving the equation of motion one can cast it
in the form
\begin{eqnarray}
 i(\gamma_{\mu}{\cal V}^{\mu})_{\alpha}
&=& -4D_{\alpha} W_{\rm eff}(\Phi), 
\label{tridentitycan}
\\
W_{\rm eff}(\Phi)&=& W(\Phi) 
+ {1\over{8}} (\bar{D}\Phi) D\Phi  
- {1\over{2}} \Phi {\delta S\over{\delta \Phi}},
\label{Weffcan}
\end{eqnarray}
with $-(1/2)\Phi\, (\delta S/\delta \Phi) = \kappa W''(\Phi)$ as
quoted  in Eq.~(\ref{phidsdphi}).
One may equally well write  $W_{\rm eff}(\Phi)$ as
\begin{eqnarray}
W_{\rm eff}(\Phi)
&=& {1\over{16}} \bar{D} D\Phi^{2} + \tilde{W}_{\rm eff}(\Phi), 
\\
\tilde{W}_{\rm eff}(\Phi)&=&
 W(\Phi) - {1\over{4}} \Phi W'(\Phi)  
-  {1\over{4}} \Phi {\delta S\over{\delta \Phi}} . 
\label{Weffimp}
\end{eqnarray}

Equation~(\ref{tridentitycan}) is a supersymmetric version of the
trace identity~\cite{CCJ},  as seen from the component expression
\begin{eqnarray}
 i(\gamma_{\mu}{\cal V}^{\mu})_{\alpha}
&=&  i(\gamma_{\mu}J^{\mu})_{\alpha} 
+2\,\theta_{\alpha}\Theta^{\mu}_{\mu}
- 2i(\gamma^{\nu}\theta)_{\alpha}\,
\epsilon_{\nu \mu \lambda}\, \zeta_{\rm can}^{\mu \lambda} 
\nonumber\\
&& +{1\over{2}}\, \bar{\theta}\theta\, (i\gamma_{\mu}f^{\mu})_{\alpha}.
\end{eqnarray}
The $(\gamma^{\nu}\theta)_{\alpha}$ component
of Eq.~(\ref{tridentitycan}), in particular, agrees with
Eq.~(\ref{zetaanom}).   This shows that the quantum modification of the
topological current $\zeta^{\mu\lambda}_{\rm can}$ derives from the
superconformal anomaly $-(1/2)\Phi\, (\delta S/\delta \Phi) = \kappa
W''(\Phi)$ in
$W_{\rm eff}(\Phi)$, as in the 2d kink case.

Let us here note that for $W(\Phi)=0$ both $\phi(x)$ and
$\psi_{\alpha}(x)$ are free and massless, and the model has exact
conformal symmetry.  Accordingly the $(1/8)\, \bar{D}\Phi D\Phi$ term
in $W_{\rm eff}(\Phi)$ is not a genuine breaking term, and it can
be removed by an appropriate redefinition, i.e., the so-called
"improvement", of the symmetry currents. 
As for the improvement~\cite{CCJ} one may recall that one is free
to modify a conserved current $j^{\mu}$ by adding a divergence of 
an antisymmetric tensor $\propto \partial_{\nu}j^{\mu\nu}$ without
changing the conservation law $\partial_{\mu}j^{\mu}=0$ and the charge
$\int d^{2}{\bf x}\,j^{0}$.

For ${\cal V}^{\mu}_{\alpha}$ 
let us try an antisymmetric spinor-tensor superfield 
${\cal I}^{\nu\mu}=\Phi\, [\gamma^{\nu},
\gamma^{\mu}]D\Phi =
-i\epsilon^{\nu\mu\lambda}\gamma_{\lambda}D\Phi^{2}$
which obeys $\partial_{\nu} (\gamma_{\mu}{\cal I}^{\nu\mu})
= -2\partial_{\nu}(
\gamma^{\nu}D\Phi^{2}) = iD_{\alpha}\bar{D}D\Phi^{2}$.
We define the improved supercurrent by
${\cal V}^{\mu}_{\alpha} - (1/4)\partial_{\nu} {\cal I}_{\alpha}^{\nu\mu}$
or
\begin{equation}
\tilde{\cal V}^{\mu}_{\alpha}
= {\cal V}^{\mu}_{\alpha} - {i\over{4}}\, \epsilon^{\mu\nu\rho}
\partial_{\nu}(\gamma_{\rho}D\Phi^{2})_{\alpha},
\label{impsfscurrent}
\end{equation}
which then satisfies the "improved" trace identity 
\begin{equation}
i(\gamma_{\mu}\tilde{\cal V}^{\mu})_{\alpha}
= -4D_{\alpha} \tilde{W}_{\rm eff}(\Phi)
\label{impTI}
\end{equation}
with the superpotential $\tilde{W}_{\rm eff}(\Phi)$
defined in Eq.~(\ref{Weffimp}).

Passing from ${\cal V}^{\mu}_{\alpha}$ to 
$\tilde{\cal V}^{\mu}_{\alpha}$ 
yields the following supermultiplet of  improved symmetry currents 
\begin{eqnarray}
\tilde{J}^{\mu}_{\alpha}&=&
J^{\mu}_{\alpha} -{i\over{2}}\, \epsilon^{\mu\nu\rho}
\partial_{\nu}\{ (\gamma_{\rho}\psi)_{\alpha}\phi\},
\nonumber\\ 
\tilde{\Theta}^{\mu \lambda} &=& \Theta^{\mu \lambda}
+{1\over{8}}\,(g^{\mu\lambda}\partial^{2}
- \partial^{\mu}\partial^{\lambda})\phi^{2},\nonumber\\ 
\tilde{\zeta}^{\mu\lambda}&=& \epsilon^{\mu \lambda\nu}\Big\{
F\partial_{\nu}\phi - {1\over{4}}\,\partial_{\nu}(F\phi) \Big\},
\nonumber\\ 
\tilde{f}^{\mu}_{\alpha}
&=&  f^{\mu}_{\alpha}
+{i\over{2}}\, (g^{\mu\nu}\partial^{2} -\partial^{\mu}\partial^{\nu})
(\gamma_{\nu}\psi\phi)_{\alpha}.
\end{eqnarray}
Interestingly, the present improvement has removed 
the antisymmetric component 
$T^{\mu \lambda}_{\rm asym} \propto 
\epsilon^{\mu \lambda\nu}\partial_{\nu}(\bar{\psi}\psi)$ from
$T^{\mu\lambda}$, yielding the symmetric energy-momentum tensor
$\tilde{\Theta}^{\mu \lambda}$ with a well-known improvement
term~\cite{CCJ} and the topological current
$\tilde{\zeta}^{\mu\lambda}$ involving no fermion field. 
From Eq.~(\ref{impTI}) follow the improved trace identity
and its superpartners:
\begin{eqnarray}
 i\gamma_{\mu}\tilde{J}^{\mu}
&=&  -4\psi\,  \tilde{W}_{\rm eff}'(\phi)
\nonumber\\ 
\tilde{\Theta}^{\mu}_{\mu}&=& 
-2 F\, \tilde{W}_{\rm eff}'(\phi)  
+ \tilde{W}_{\rm eff}''(\phi)\, \bar{\psi}\psi
\nonumber\\ 
\tilde{\zeta}^{\mu \lambda} 
&=&  -\epsilon^{\mu \lambda\nu}\, \partial_{\nu}\tilde{W}_{\rm eff}(\phi),
\nonumber\\ 
 i\gamma_{\mu}\tilde{f}^{\mu}
&=& -4i \partial_{\mu} [\gamma^{\mu}\psi \tilde{W}_{\rm eff}'(\phi)] .
\label{impTIcomponent}
\end{eqnarray}

With $\tilde{\cal V}^{\mu}_{\alpha}$  one again finds a conserved-charge
superfield
\begin{equation}
\int_{-\infty}^{\infty} d^{2}x \tilde{\cal V}^{0}_{\alpha}
= \tilde{Q}_{\alpha} -2i(\gamma_{\lambda}\theta)_{\alpha} 
(\tilde{P}^{\lambda} + \tilde{Z}^{\lambda}) ,
\label{impsfcharge}
\end{equation}
which shows that the improved charges
$\tilde{Q}_{\alpha}, \tilde{P}^{\lambda}$ and $\tilde{Z}^{\lambda}$ 
obey the same supercharge algebra as in Eq.~(\ref{superchargealgebra}). 
Note that $\tilde{Q}_{\alpha}$ and $\tilde{P}^{\lambda}$ are essentially
the same as the original charges $Q_{\alpha}$ and $P^{\lambda}$, 
under the same asymptotic $({\bf x} \rightarrow \pm \infty)$ 
condition on the fields as discussed for the fermionic charge 
$\int d^{2}{\bf x}\, f^{0}_{\alpha}$ in Eq.~(\ref{fmualpha}). 
This in turn implies that the central charges 
$Z^{\lambda} = \int d^{2}{\bf x}\, \zeta^{0 \lambda}_{\rm can}$ and 
$\tilde{Z}^{\lambda} = \int d^{2}{\bf x}\, \tilde{\zeta}^{0 \lambda}$,
though different in form, are physically equivalent.

It is enlightening to verify this equivalence.
It is a simple task to evaluate, using Eqs.~(\ref{chipropagator}), 
(\ref{phidsdphi}) and~(\ref{DchiDchi}), the expectation values of the
effective superpotentials $W_{\rm eff}(\Phi)$ and 
$\tilde{W}_{\rm eff}(\Phi)$ to one loop or $O(\hbar^{0})$:
\begin{eqnarray}
\langle W_{\rm eff}(\Phi) \rangle
&=&  U_{\rm eff}(\Phi_{\rm c}) 
+ {1\over{8}} \bar{D}\Phi_{\rm c} D\Phi_{\rm c},
\\
\langle \tilde{W}_{\rm eff}(\Phi) \rangle
&=& U_{\rm eff}(\Phi_{\rm c}) 
- {1\over{4}}\, \Phi_{\rm c}\, U'_{\rm eff}(\Phi_{\rm c}),
\end{eqnarray}
apart from terms of $O(D^{2})$, 
with $U_{\rm eff}(\Phi_{\rm c})$ defined in Eq.~(\ref{Ueff}). 
Here we see that 
$\langle W_{\rm eff}(\Phi) \rangle$ and 
$\langle \tilde{W}_{\rm eff}(\Phi) \rangle$
precisely agree with $U_{\rm eff}(\Phi_{\rm c})$ 
at spatial infinities $x^{1}\rightarrow  \pm \infty$ 
where $\psi_{\rm c} \rightarrow 0$ and $\phi_{\rm c} \rightarrow \pm v$
with $v$ determined from $U'_{\rm eff}(v)=0$. 
The resulting  central charges
\begin{equation}
\langle Z_{\rm can}^{\lambda}\rangle/L_{y} =
\langle \tilde{Z}^{\lambda}\rangle/L_{y} 
= \delta^{\lambda 2} 2 U_{\rm eff}(v)
\end{equation}
are thus in agreement with Eq.~(\ref{dwtension}) obtained from the
effective action.

\section{Induced spin and extended supersymmetry}

In this section we explore physical origins of the central charge anomaly.
In Eq.~(\ref{psibarpsi}) we have seen that the main
quantum correction to the central charge comes from 
the ${1\over{8}} \langle \bar{\psi}\psi \rangle$ portion of 
$\langle Z^{2} \rangle$.
In three dimensions, with the spatial-rotation matrix 
$\sigma^{12} \equiv (i/2)\, [\gamma^{1},\gamma^{2}]=
\sigma_{2}=\gamma^{0}$,
the fermion mass term is nothing but the spin density 
$\psi^{\dag}{1\over{2}}\,\sigma^{12}\psi ={1\over{2}}\,\bar{\psi}\psi$.
Accordingly, through the $\lq\lq$Zeemann coupling" $W''\bar{\psi}\psi$,
the domain-wall configuration works to align the fermion spin oppositely
in the two domains; e.g., for $x^{1}<0$, $W'' > 0$ so that  
$\langle \bar{\psi}\psi\rangle < 0$ is preferred.
The quantum central charge 
$\sim \langle \bar{\psi}\psi \rangle$ is therefore ascribed to 
induced spin in the domain-wall background.

Actually, it is possible to evaluate the induced polarization
reliably with free fermions if one notes that, except for the vicinity 
of the wall, the effective fermion mass is almost
constant $W''\approx \mp m$ in each domain.
Consider the relativistic expression 
$\langle \bar{\psi}(x)\, \psi (x)\rangle 
= -{\rm tr}\langle x|i/({\slash \!\!\!p} -m)|x\rangle$ 
and integrate over $p^{0}$ first. 
The result clarifies the meaning of $\langle \bar{\psi}\psi \rangle$: 
It is written as a sum over (twice) aligned spins of negative-energy
fermions filling the Dirac sea, the fermionic vacuum:
\begin{equation}
\langle \bar{\psi} \psi\rangle 
= - \sum_{\bf p} {m\over{|m|}} 
= - {m\over{2\pi}} \Big(\sqrt{ \Lambda^{2}+m^{2}} -|m|\Big),
\end{equation}
where  $\sum_{\bf p} \equiv 
\int \{d^{2}{\bf p}/(2\pi)^{2}\} |m|/\epsilon 
=|m| \int d\epsilon/(2\pi)$
is the phase-space sum 
and $\epsilon=\sqrt{m^{2}+ {\bf p}^{2}}$.
The divergent contribution involving the momentum cutoff $\Lambda^{2}$
is associated with the infinite depth 
or infinite phase space of the Dirac sea~\cite{fninducedspin}.  The
conformal anomaly $\sim \kappa W''(\phi)$ works to cancel this
infinite  intrinsic spin of the fermionic vacuum, 
leaving finite induced spin $m^{2}/(2\pi)$ for the central charge, 
as we have seen in Eq.~(\ref{psibarpsi}).

Note here that, upon improvement~(\ref{impsfscurrent}), 
the induced fermion spin 
$\sim {1\over{8}}\langle \bar{\psi}\psi\rangle$ 
in the topological current
$\zeta^{\mu\lambda}_{\rm can}$ is
transferred into an induced (bosonic) superpotential 
$- {1\over{4}} \langle \phi W'(\phi) \rangle + {1\over{2}} \kappa 
\langle W''(\phi)\rangle$  in the improved current 
$\tilde{\zeta}^{\mu\lambda}$; see Eq.~(\ref{impTIcomponent}).  
This reveals a dual (fermionic/bosonic) character of the central-charge
anomaly. This dualism is unexpected but is quite natural since in
supersymmetric theories  fermionic and bosonic quantum fluctuations are
intimately related, as seen from Eq.~(\ref{DchiDchi}).

The dual character of the anomaly can be verified for 
the 2d kink case~\cite{KScc} as well.
There the trace identity is governed by the superpotential 
\begin{equation}
W^{\rm kink}_{\rm eff}(\Phi)
= W(\Phi) - {1\over{2}}\,\Phi {\delta S\over{\delta \Phi}}
\end{equation}
and the central-charge anomaly comes from the superconformal anomaly
$-{1\over{2}}\,\Phi (\delta S/\delta \Phi) = W''(\Phi)/(4\pi)$.
We try the following improvement. 
Let ${\cal I}_{2}^{\nu\mu}= [\gamma^{\nu}, \gamma^{\mu}]D\Phi^{2}$ 
and consider the improved supercurrent 
$\tilde{\cal V}^{\mu}_{\alpha} = {\cal V}^{\mu}_{\alpha} 
- (1/4)\partial_{\nu} ({\cal I}_{2})_{\alpha}^{\nu\mu}$. 
It satisfies the "improved" trace identity 
\begin{equation}
i(\gamma_{\mu}\tilde{\cal V}^{\mu})_{\alpha}
= -2D_{\alpha}\tilde{W}^{\rm kink}_{\rm eff}(\Phi)
\end{equation}
with the new effective superpotential
\begin{equation}
\tilde{W}^{\rm kink}_{\rm eff}(\Phi) 
= W(\Phi) - {1\over{2}} \Phi W'(\Phi) 
- {1\over{4}} (\bar{D}\Phi) D\Phi.
\label{Wefftwodim}
\end{equation}
Here the improvement has been made to remove the anomaly 
from $W^{\rm kink}_{\rm eff}(\Phi)$.
One can then verify that the central-charge
anomaly entirely resides in the fermion sector 
$- {1\over{4}} \langle \bar{D}\Phi\, D\Phi \rangle 
= W''(\Phi_{\rm c})/(4 \pi) + C$ 
while 
$- {1\over{2}}\, \langle \Phi W'(\Phi) \rangle 
= - {1\over{2}}\Phi_{\rm c}\, U'_{\rm eff}(\Phi_{\rm c}) - C$
works to remove the divergent piece 
$C \sim (W''_{\rm c}/8\pi) \log [\Lambda^{2}/(W''_{\rm c})^{2}]$ 
from the former. 
Thus in this case one would interpret the central-charge anomaly as 
due to induced quantum number $\langle \bar{\psi} \psi\rangle$  
in the kink background.

Finally, as for the presence or absence of anomalies in the central
charge it is instructive to look into the case of $N=2$ supersymmetry, 
for which, in 1+1 dimensions, the central-charge anomaly is known to be
absent~\cite{NSNR,SVV}.  The relevant $N=2$ model in (two or) three
dimensions is obtained from the 4d Wess-Zumino model 
via dimensional reduction.  
In terms of two $N=1$ real superfields
$\Phi_{1}(z)$ and
$\Phi_{2}(z)$ the superspace action is expressed in the
form~(\ref{sfaction}) with the kinetic term  
$\bar{D}\Phi D\Phi \rightarrow \sum_{i}\bar{D}\Phi_{i}D\Phi_{i}$
and the superpotential~\cite{SVV}
\begin{equation}
W(\Phi_{1}, \Phi_{2})= {m^{2}\over{4\lambda}}\, \Phi_{1} 
- {\lambda\over{3}}\, \Phi_{1}^{3} + \lambda \Phi_{1}\Phi_{2}^{2},
\label{NtwoWZmodel}
\end{equation}
which is harmonic, $\sum_{i}W_{ii} =0$
with  $W_{ij} \equiv \partial^{2}W/\partial\Phi_{i}\partial\Phi_{j}$,
a property characteristic of extended supersymmetry. The classical
domain-wall configuration is realized with
$\phi_{1}(x) \rightarrow \phi_{\rm DW}(x)$ and
$\phi_{2}(x) \rightarrow 0$.

The conserved supercurrent is written as a sum 
${\cal V}^{\mu}_{\alpha}= {\cal V}^{\mu}_{\alpha}[\Phi_{1}] +
{\cal V}^{\mu}_{\alpha}[\Phi_{2}]$ of
${\cal V}^{\mu}_{\alpha}$ in Eq.~(\ref{sfcurrent}) 
and the associated trace identity is written as 
$i(\gamma_{\mu}{\cal V}^{\mu})_{\alpha}= -4D_{\alpha} W_{\rm eff}$ 
with
\begin{equation}
W_{\rm eff}= W(\Phi_{1},
\Phi_{2})  + {1\over{8}}(\bar{D}\Phi_{i})
D\Phi_{i} - {1\over{2}}\,
\Phi_{i} (\delta S/\delta \Phi_{i}).
\end{equation}
The central charge is now read from this $W_{\rm eff}$.
Note that in each domain the two species of fermions $D\Phi_{1,2}$
have effective masses opposite in sign, 
$W_{11}(\phi_{i}) = - W_{22}(\phi_{i}) = -2\lambda \phi_{1} \sim -m$.
The induced spin $\langle \bar{D}\Phi_{i} D\Phi_{i} \rangle$
from each species therefore differs in sign, 
yielding no net polarization in each domain. 
No infinite polarization or no intrinsic short-distance anomaly
thereby remains with the sum 
$\sum_{i} \Phi_{i} (\delta S/\delta \Phi_{i}) \sim \sum_{i}W_{ii}$
vanishing, leaving no anomaly in the central charge.
The absence of induced spin and that of the central-charge anomaly are
thus correlated. This is not a coincidence. It is a consequence of the
nonrenormalization theorem~\cite{WB,NonRenTh} which states that there is
no quantum correction to chiral superpotentials. 
The chiral structure inherent in $N=1$ supersymmetry 
in four dimensions is responsible for the absence of 
the central-charge anomaly in the present dimensionally-reduced model
with $N=2$ supersymmetry.

\section{Summary and discussion}

In the present paper we have studied some general features of 
the central-charge anomaly for domain walls in three-dimensional
supersymmetric theories.  
The way the anomaly arises in the supercharge algebra
critically depends on the dimension of spacetime.
For kinks in two dimensions the central-charge anomaly
arises as part of the superconformal anomaly.
For domain walls in three dimensions the central charge has, 
besides the superpotential $W(\phi)$, a fermion-spin term 
$\sim \bar{\psi}\psi$ at the classical level.  
The quantum modifications of the supercharge algebra therefore come
not only from the short-distance anomaly but also
from quantum induced spin $\sim \langle \bar{\psi}\psi \rangle$, and
the latter dominates in the sum.  
For domain walls the central-charge anomaly is thus ascribed to quantum
induced spin of long-distance origin.

The best place to explore the central-charge anomaly is the supersymmetric
trace identity, in view of the fact that the topological
current lies in a supermultiplet together with the energy-momentum tensor
and supercurrent.
This naturally has led us to consider the improvement of the superfield
supercurrent (since one normally has to improve the canonical
energy-momentum tensor to arrive at the well-behaved conformal
currents~\cite{CCJ}).
We have thereby seen that the anomaly in the central charge, upon
improvement, can be transferred from induced spin in the fermion sector
to an induced potential in the boson sector, or vice versa.  This
has revealed an unexpected dual character, both fermionic and bosonic, of
the central-charge anomaly.  
This (boson/fermion) dualism has a further consequence for kinks
in two dimensions.  There one can make an improvement so that
the short-distance anomaly is transformed into induced fermion
quantum number $\sim \langle \bar{\psi}\psi \rangle$; the
central-charge anomaly thus has a dual character of either short- or
long-distance origin as well.

We have also examined the case of extended supersymmetry and noted
that  the absence or presence of the short-distance anomaly and that of
induced spin are correlated.  This coincidence is quite natural in the
light of the dual character of the central-charge anomaly, which itself
is a reflection of the underlying supersymmetry.

Finally it would be worth remarking that 
extensive use of the superfield supercurrent is crucial to our analysis
of the central-charge anomaly.
It makes manifest the supermultiplet nature of various symmetry
currents, conservation laws and the associated anomalies. 
Use of superfields has also helped us systematize the process of
improvement of the supercurrent, which, if done separately for
each component current, could have been a laborious task.

\acknowledgments

This work was supported in part by a Grant-in-Aid for Scientific Research
from the Ministry of Education of Japan, Science and Culture (Grant No.
14540261).


\begin{thebibliography}{99}
\bibitem{fninducedspin}
A result consistent with Eq.~(\ref{DchiDchi}) is obtained if one evaluates
$\langle \bar{\psi} \psi\rangle$ 
using the covariant regulator $\exp[\tau (\slash\!\!\! p -m)^{2}]$.
\end{thebibliography}
\end{document}